\documentclass[a4paper]{jpconf}
\usepackage{graphicx}
\usepackage{bm}
\usepackage{amsmath}
\usepackage{amsfonts}
\usepackage{dutchcal}
\usepackage{hyperref}
\usepackage{xcolor}
\usepackage{todonotes}

\hypersetup{
    colorlinks,
    linkcolor={cyan!50!black},
    citecolor={magenta!70!black},
    urlcolor={magenta!80!black}
}
\DeclareMathAlphabet{\mathdutchcal}{U}{dutchcal}{m}{n}

\newcommand{\unvec}[1]{\bm{\hat{#1}}}
\newcommand{\pow}[2]{#1^{(#2)}}
\newcommand{\dd}{\mathrm{d}}
\newcommand{\sech}{\mathrm{sech}}
\newcommand{\tensor}[1]{\stackrel{\leftrightarrow}{#1}}
\begin{document}

\title{Non-linear effective field theory simulators in two-fluid interfaces}

\author{Vitor S. Barroso$^{1,2,3}$, Cameron R. D. Bunney$^{1,2,3}$, Silke Weinfurtner$^{1,2,3}$}
\address{$^{1}$ School of Mathematical Sciences, University of Nottingham, University Park, Nottingham, NG7 2RD, UK}
\address{$^{2}$Centre for the Mathematics and Theoretical Physics of Quantum Non-Equilibrium Systems, University of Nottingham, Nottingham, NG7 2RD, UK}
\address{$^{3}$Nottingham Centre of Gravity, University of Nottingham, Nottingham, NG7 2RD, UK}
\ead{vitor.barrososilveira@nottingham.ac.uk}
\ead{cameron.bunney@nottingham.ac.uk}
\ead{silke.weinfurtner@nottingham.ac.uk}


\begin{abstract}
Analogue gravity offers an approach for testing the universality and robustness of quantum field theories in curved spacetimes and validating them using down-to-earth, laboratory-based experiments. Fluid interfaces are a promising framework for creating these gravity simulators and have successfully replicated phenomena such as Hawking radiation and black hole superradiance. Recent work has shown that hydrodynamical instabilities on the interface between two fluids can capture features of the post-inflationary thermalisation of the Early Universe. In this study, we extend fluid dynamics methods to develop an effective field theory for the interface between two fluids, demonstrating the equivalence between the governing equations and a relativistic scalar field in an analogue spacetime. We also show that the interfacial height field serves as the analogue relativistic field even in a nonlinear, interacting field theory. We propose that these mathematical equivalences can be extrapolated to probe regimes where calculations are challenging or impractical.  Our work provides a new framework for simulating far-from-equilibrium cosmological and gravitational scenarios in the laboratory.
\end{abstract}

\section{Introduction}
Over four decades ago, William Unruh demonstrated that the arguments leading to Hawking radiation also predict a thermal spectrum of sound waves from the sonic horizon in transonic fluid flows~\cite{1981UnruhAnalogueGravity}, from which arose analogue gravity. This field now resonates with several areas throughout physics~\cite{barcelo2011analogue,jacquet2020next}, with experimental and theoretical efforts in ultracold atoms systems~\cite{Gooding:2020scc,Unruh:2022gso}, superfluids~\cite{bunney2023third}, and optical systems~\cite{Belgiorno2010HawkingFilaments,Philbin2008Fiber-opticalHorizon}. Many of these gravity simulators rely on mathematical equivalences between the dynamical equations of quantum fields in curved spacetimes and an Effective Field Theory (EFT) emerging from the underlying physical description of the simulator at hand. Based on these correspondences, analogue gravity leans on the notions of universality, robustness and the validation of the phenomena that it wishes to probe~\cite{jacquet2020next}. 

Fluid interfaces have been a successful platform for the experimental realisation of various gravity simulators, see e.g.~\cite{Rousseaux2008ObservationEffect,Weinfurtner2011MeasurementSystem, Torres2017RotationalFlow,Torres20QNM}. Their accessibility and broad applicability with a wide range of physical properties and flows enable the creation and manipulation of multiple analogue scenarios. However, we note that these systems exploit the linear dynamics of long-wavelength surface waves in the so-called shallow-water regime, where mathematical analogies are guaranteed~\cite{schutzhold2002gravity}. On the other hand, in recent work~\cite{barroso2022primary}, we showed that the non-linear evolution of hydrodynamical instabilities on the interface between two fluids captures dynamical features of post-inflationary thermalisation of the Early-Universe. Our experiments open a path towards using analogue systems for the investigation of interacting field theories and far-from-equilibrium scenarios while accounting for general features of the simulator, such as dispersion and damping. 

In the present work, we provide an extension of pre-existing methods in fluid dynamics to describe an EFT for the interface between two fluids~\cite{barroso2022primary}. In \Sref{sec::interfacial dynamics}, we begin by providing the fluid dynamical description of a gravity simulator, comprising of two immiscible, incompressible fluids of different densities with negligible viscosities confined within a basin, which is itself subject to a vertical mechanical forcing. The fluid of lower density, therefore, rests above the other, forming a dynamic interface due to the forcing. The interfacial dynamics form the basis for our gravity simulator with the interfacial height field $\xi$ as the analogue relativistic field.

We move then to provide two further descriptions: a variational approach and a Lagrangian formulation. The fluid dynamical equations and their boundary conditions governing gravity surface waves constitute a kinematic boundary-value problem~\cite{Miles1976NonlinearBasins,serrin1959mathematical}, which may be derived by the variation of a certain action functional. We show that this also holds true for the interfacial dynamics in our gravity simulator. In our second description, by considering the kinetic and potential energies of the fluids, we construct a Lagrangian through the spectral decompositions of the analogue relativistic field and its conjugate momentum.

Having considered and extended the general formalism to two-fluid interfaces, we exemplify in \Sref{sec:simulators} limiting forms of the governing equations, which give rise to equivalences to a relativistic scalar field in an analogue spacetime. In line with the work of Sch{\"u}tzhold and Unruh~\cite{schutzhold2002gravity} and the origins of analogue gravity, we show that in the limit of long-wavelength surface waves without external forcing, the equations of motion for the height field reduce to the dynamics of a massless Klein-Gordon field in Minkowski spacetime with a linear dispersion relation. 

We then turn our attention to cosmological simulators. By allowing again an arbitrary mechanical forcing in the system, generating an effective gravity, we consider the linearised equations of motion at the interface. After a brief review of the core equations in cosmology, the analogue reveals itself in the equations of motion for the fluid velocity potential, i.e., the conjugate momentum of the interfacial height field. These equations of motion describe a massless scalar field in a $(2+1)-$dimensional expanding Universe. The height field, however, is still an important component if one specialises in the case of harmonic forcing. The equations of motion for the height field are then the Mathieu equation, whose set of solutions are exponentially unstable in time. The Mathieu equation is cosmologically relevant in models of thermalisation in the Early Universe, and thus, the interfacial height field serves as the analogue relativistic field even in a nonlinear, interacting field theory.

\section{Non-linear interfacial dynamics between two fluids}\label{sec::interfacial dynamics}

Our starting point is a description of the emerging dynamics at the interface between two fluids enclosed in a basin subject to an external acceleration. For convenience, we consider a series of idealised assumptions on the physical properties of the fluids and later discuss the implications of more realistic traits to our derived model. Both fluid phases are taken to be immiscible, incompressible and with negligible viscosities. The geometry of the fluid cell is such that a vertical coordinate $z$ aligned with Earth's gravitational acceleration $\bm{g_0}=-g_0\unvec{z}$ is always perpendicular to a set of two-dimensional coordinates $\bm{x}$ mapping the horizontal cross-section $\Sigma$ of the cell. 

In the presence of the external forcing $\bm{F_0}(t)=F(t)\unvec{z}$, the fluids experience a combined vertical acceleration $\bm{g}(t)=-(g_0-F(t))\unvec{z}\equiv-g(t)\unvec{z}$, and waves may appear at the interface. Thus, we assume that the average interface position is at $z=0$ and parametrise interfacial waves $\xi(t,\bm{x})$ by a surface $\Gamma(t,\bm{x},z) = z - \xi(t,\bm{x})=0$. The denser fluid phase is vertically confined by the basin's bottom lid at $z=h_1=-|h_1|$ and has uniform density $\rho_1$. Conversely, the lighter phase with uniform density $\rho_2<\rho_1$ is vertically confined by the basin's top lid at $z=h_2$. The volume of each fluid is then $V_j=|h_j|\Sigma$.

In the absence of a background flow, incompressibility requires that the velocity $\bm{u}_j$ of fluid $j$ is divergenceless, in accordance with the continuity equation. We assume that the fluid flow is irrotational $(\nabla\times\bm{u}_j=\bm{0})$ and is therefore described by a potential $\phi_j$, such that $\bm{u}_j=\nabla\phi_j$ and satisfying
\begin{equation}
    \label{eq:Laplace}
    \nabla^2\phi_j=0,\text{ in }V_j.
\end{equation}
A condition of impenetrability on all rigid boundaries $\partial V_j$ with outwardly directed normal unit vectors $\unvec{n}$ then reads
\begin{equation}
    \label{eq:nopen}
    \left.\unvec{n}\cdot \nabla\phi_j\right|_{\partial V_j}=0.
\end{equation}

A kinematic boundary condition is also in place at the moving interface. It requires that fluid particles on the interface $\Gamma=0$ remain part of it and move along the flow, i.e., the interface is advected by the fluids and~\cite{Landau2013Fluid6,serrin1959mathematical}
\begin{equation}
    \label{eq:kinBC}
    \frac{D\Gamma}{Dt}\equiv \frac{\partial \Gamma}{\partial t}+\nabla\phi_j\cdot\nabla\Gamma=0,\text{ on }\Gamma=0.
\end{equation}
We further stress that the equation above constitutes two equations, one in each fluid, and expands to
\begin{equation}
    \label{eq:kinematic}
    \frac{\partial\xi}{\partial t} = \frac{\partial\phi_j}{\partial z}-\nabla\phi_j\cdot\nabla\xi, \text{ on }z=\xi(t,\bm{x}),
\end{equation}
by noting that $\nabla\Gamma=\unvec{z}-\nabla\xi$.
It is now clear that the equation above, to a linear approximation, requires that the time evolution of the interface equals the vertical velocity component $u_{j,z}=\partial_z\phi_j$ of both fluids.

Under the assumptions presented here, the bulk motion of each fluid is governed by Euler's equations for a potential flow, which we express in the form of a conservation condition, as follows,
\begin{equation}
\label{eq:eom}
    \rho_j\left[\frac{\partial}{\partial t}(\nabla\phi_j)+\nabla\phi_j\cdot\nabla\left(\nabla\phi_j\right)\right]=\nabla\cdot\bm{\tensor{\pi}}_j+\bm{f}_j,
\end{equation}
where $\bm{\tensor{\pi}}_j$ and $\bm{f}_j$ are the stress tensor and the external force density on fluid $j$~\cite{Landau2013Fluid6,Kumar94}. For negligible viscosities, the components $\pi_{j,kl}$ of the stress tensor are given by
\begin{equation}
\label{eq:stressTensor}
    \pi_{j,kl}=-p_j\delta_{kl},
\end{equation}
where $p_j$ is the pressure in fluid $j$, and the indices $k,l$ refer to the $\bm{x},z$ components of $\bm{\tensor{\pi}}_j$. The combined acceleration $g(t)$ enters equation~\eqref{eq:eom} upon identifying $\bm{f}_j=-\rho_j g(t)\unvec{z}$. We can then join equation~\eqref{eq:eom} of each fluid, evaluate it at the interface and rearrange terms to obtain
\begin{equation}
\label{eq:eom12}
    \left[\frac{\partial}{\partial t}\left(\rho_1\phi_1-\rho_2\phi_2\right)+\frac{\rho_1}{2}\left(\nabla\phi_1\right)^2-\frac{\rho_2}{2}\left(\nabla\phi_2\right)^2\right]_{z=\xi}
    =\left(\left.\pi_{1,zz}\right|_{z=\xi^-}-\left.\pi_{2,zz}\right|_{z=\xi^+}\right)-(\rho_1-\rho_2)g(t)\xi,
\end{equation}
where the first term on the right-hand side indicates the jump of the vertical component of stress across the interface.

When the two fluid phases are in motion, it is often the case that interfacial-tension forces must be taken into account. To incorporate these effects into the equations of motion, we employ the Young-Laplace law, which states that the stress across the interface is proportional to its surface curvature~\cite{Landau2013Fluid6}, i.e.,
\begin{equation}
    \label{eq:YoungLaplace}
    \left(\left.\pi_{1,zz}\right|_{z=\xi^-}-\left.\pi_{2,zz}\right|_{z=\xi^+}\right)=-\sigma\nabla\cdot\unvec{n}_\Gamma=-\sigma\nabla\cdot\left(\frac{\nabla\Gamma}{|\nabla\Gamma|}\right)=\sigma\nabla\cdot\left(\frac{\nabla\xi}{\sqrt{1+|\nabla\xi|^2}}\right),
\end{equation}
where $\sigma$ denotes the interfacial-tension coefficient between the two fluids. For small surface gradients, $|\nabla\xi|\ll 1$, the equation above recovers the usual surface tension term $-\sigma\nabla^2\xi$. Finally, the resulting interfacial dynamics are determined by the kinematic boundary condition~\eqref{eq:kinematic} and the following equation of motion,
\begin{equation}
\label{eq:interfacialDyn}
    \left[\frac{\partial}{\partial t}\left(\rho_1\phi_1-\rho_2\phi_2\right)+\frac{\rho_1}{2}\left(\nabla\phi_1\right)^2-\frac{\rho_2}{2}\left(\nabla\phi_2\right)^2\right]_{z=\xi}
    +(\rho_1-\rho_2)g(t)\xi
    -\sigma\nabla\cdot\left(\frac{\nabla\xi}{\sqrt{1+|\nabla\xi|^2}}\right)=0.
\end{equation}

\subsection{Variational approach to the kinematic boundary-value problem}

In line with~\cite{Miles1976NonlinearBasins}, we note that equations~\eqref{eq:Laplace},~\eqref{eq:nopen} and~\eqref{eq:kinematic} constitute a kinematic boundary-value problem and may be derived from the stationarity of an action functional~\cite{serrin1959mathematical} given by
\begin{equation}
    \label{eq:actionDirichlet}
    I_j = \frac{(-1)^{j+1}}{2\Sigma}\iiint_{V_j} \left(\nabla\phi_j\right)^2 \dd\Sigma~\dd z
    -\frac{1}{\Sigma}\iint \frac{1}{|\nabla\Gamma|}\frac{\partial\xi}{\partial t}\left.\phi_j\right|_{z=\xi} \dd\Sigma,
\end{equation}
where $\dd\Sigma$ is the area element of the horizontal cross-section and is such that $\Sigma = \iint\dd\Sigma$. In~\ref{app:variationAction}, we show that the variation of action $I_j$ with respect to $\phi_j$ recovers the appropriate set of equations\footnote{The second integral in equation~\eqref{eq:actionDirichlet} diverges in form from that presented in~\cite{Miles1976NonlinearBasins}, but it is in accordance with the Dirichlet problem as presented in Serrin~\cite{serrin1959mathematical} and recovers equation~\eqref{eq:kinematic} without the need of any approximations or perturbative expansions.}. The sign in front of the first integral is necessary to recover equation~\eqref{eq:kinematic} appropriately in fluid $2$, where the normal at the interface is opposite to that of fluid $1$\footnote{This sign change did not appear in Miles~\cite{Miles1976NonlinearBasins} as the author was considering scenarios where only fluid $1$ is present.}.

We now assume that the velocity potentials may be decomposed into time-dependent generalised coordinates $\phi_{j,a}(t)$ associated with a discrete set of spatial eigenfunctions $\psi_{j,a}(\bm{x},z)$ of Laplace's equation~\eqref{eq:Laplace} with boundary conditions~\eqref{eq:nopen}, i.e.,
\begin{equation}
    \label{eq:phidecomp}
    \phi_j(t,\bm{x},z) = \sum_a \phi_{j,a}(t)\psi_{j,a}(\bm{x},z).
\end{equation}
The symmetry required by the fluid cell geometry allows us to decouple the $z-$dependence of each fluid from the horizontal component common to both. Hence, we employ the decomposition $\psi_{j,a}(\bm{x},z)=\chi_a(\bm{x})f_{j,a}(z)$, with 
\begin{equation}
    \label{eq:z-component}
    f_{j,a}(z)=\frac{\cosh[k_a(z-h_j)]}{\cosh(k_ah_j)},
\end{equation}
for $k_a$ such that\footnote{In equation~\eqref{eq:horizontalHelm}, it is left implicit that $\nabla\equiv\nabla_{\bm{x}}$.}
\begin{equation}
    \label{eq:horizontalHelm}
    \left(\nabla^2+k_a^2\right)\chi_a(\bm{x})=0,\text{ with }\left.\unvec{n}\cdot\nabla\chi_a\right|_{\partial \Sigma}=0.
\end{equation}
The function in~\eqref{eq:z-component} is constructed to identically satisfy the boundary conditions at the rigid top and bottom lids of the cell, namely $\partial_z \phi_j(t,\bm{x},z=h_j)=0$. We also require that the set of functions $\chi_a$ is orthonormal on the cross-section $\Sigma$, i.e.,
\begin{equation}
    \label{eq:normalisation}
    \iint\dd\Sigma~\chi_a(\bm{x})\chi_b(\bm{x}) = \Sigma \delta_{ab}.
\end{equation}
Finally, we suppose that a similar decomposition to~\eqref{eq:phidecomp} applies to the height fluctuations $\xi(t,\bm{x})$, as follows
\begin{equation}
    \label{eq:xidecomp}
    \xi(t,\bm{x})=\sum_{a} \xi_{a}(t)\chi_{a}(\bm{x}). 
\end{equation}

In the basis of eigenfunctions $\chi_a$, we can express the generalised coordinates $\xi_a$ and $\phi_{j,a}$ as columns vectors of infinite dimension, namely $\bm{\Phi}_j\equiv\{\phi_{j,a}\}$ and $\bm{\Xi}\equiv\{\xi_{a}\}$, and similarly for the basis, $\bm{X}\equiv\{\chi_a\}$. With this notation, equations~\eqref{eq:phidecomp} and~\eqref{eq:xidecomp} may be written as a dot product of vectors, e.g., $\xi(t,\bm{x})=\bm{\Xi}^\mathrm{T}\bm{X}$, where $\bm{\Xi}^\mathrm{T}$ denotes the transpose of $\bm{\Xi}$. Then, the action functional in~\eqref{eq:actionDirichlet} reads
\begin{equation}
\label{eq:action}
    I_j = \frac{1}{2}\bm{\Phi}_j^\mathrm{T}\bm{\mathrm{K}}_j\bm{\Phi}_j-\frac{1}{2}\bm{\dot{\Xi}}^\mathrm{T}\bm{\mathrm{D}}_j\bm{\Phi}_j,
\end{equation}
where $\bm{\mathrm{K}}_j\equiv\left[\pow{\mathdutchcal{k}}{j}_{ab}\right]$ and $\bm{\mathrm{D}}_j\equiv\left[\pow{\mathdutchcal{d}}{j}_{ab}\right]$ are two square matrices defined in terms of integrals of products of $\chi_a$, $\psi_a$ and $\xi_a$, given explicitly in~\ref{app:MatrixInt}. In deriving these terms, we assume that changes in the interfacial height $\xi(t,\bm{x})$ are much smaller than the depth of both fluids, i.e., $|\xi|\ll |h_j|$. In practice, we suppose that the appropriate quantities can be perturbatively expanded for $\left|\tfrac{\xi}{h_j}\right|\ll 1$ with contributions up to second-order terms.

By varying the action $I_j$ with respect to $\bm{\Phi}_j$ and noting that $\bm{\mathrm{K}}_j$ is symmetric (see~\ref{app:MatrixInt}), we obtain the corresponding kinematic boundary-value problem in matrix notation, as follows,
\begin{equation}
    \label{eq:kinematicMatrix}
    \bm{\Phi}_j = \bm{\mathrm{K}}_j^{-1}\bm{\mathrm{D}}_j^{\mathrm{T}}\bm{\dot{\Xi}}\equiv \bm{\mathrm{L}}_j\bm{\dot{\Xi}},
\end{equation}
where $\bm{\mathrm{K}}_j^{-1}$ denotes the inverse of $\bm{\mathrm{K}}_j$ and $\bm{\mathrm{D}}_j^{\mathrm{T}}$ is the transpose of $\bm{\mathrm{D}}_j$. The exact form of the combined square matrix $\bm{\mathrm{L}}_j\equiv\left[\pow{\mathdutchcal{l}}{j}_{ab}\right]$ is given in equation~\eqref{eq:app:lmatrix} of~\ref{app:MatrixInt} along with its derivation. The equation above to leading order approximation recovers the linearised dynamics result (cf.~\cite{Kumar94}), i.e., 
\begin{equation}
\label{eq:linearKinematic}
    \dot{\xi}_a = k_a\tanh(k_a |h_1|)\phi_{1,a}=-k_a\tanh(k_a |h_2|)\phi_{2,a}.
\end{equation}
Equation~\eqref{eq:kinematicMatrix} provides a mapping between the velocity potential modes $\phi_{j,a}$ and the height velocity modes $\dot{\xi}_{a}$, and the dependence of $\pow{\mathdutchcal{l}}{j}_{ab}$ on the modes $\xi_a$ results in the nonlinear generalisation of equation~\eqref{eq:linearKinematic}.

\subsection{Lagrangian formulation}

The set of equations given by~\eqref{eq:kinematic} and~\eqref{eq:interfacialDyn} determines the dynamics of the interfacial modes between the two fluids. A common approach to recovering these equations of motion is Luke's variational principle~\cite{luke1967variational}, where a Lagrangian is obtained by integrating the fluid pressures $p_j$ over their volumes $V_j$. Conversely, we opt for the approach of Miles in~\cite{Miles1976NonlinearBasins,Miles1984NonlinearResonance}, where the kinetic and potential energies of the fluids at play are considered. A Lagrangian can then be constructed from their difference using the spectral decomposition presented in the previous section (see equations~\eqref{eq:phidecomp} and~\eqref{eq:xidecomp}). The results of this approach should be in line with a decomposition of equation~\eqref{eq:interfacialDyn} in terms of interfacial modes $\xi_a$.

We begin by examining the kinetic energy $T_j$ of fluid $j$~\cite{Landau2013Fluid6}, and employing relation~\eqref{eq:kinematicMatrix} to express it in terms of the interfacial height modes $\xi_a$, as follows,
\begin{equation}
    \label{eq:kineticEnergy}
    \frac{T_j}{\Sigma} = \frac{\rho_j}{2\Sigma}\iiint_{V_j}\dd\Sigma\dd z~\left(\nabla\phi_j\right)^2
    =\frac{(-1)^{j+1}}{2}\rho_j\bm{\Phi}_j^\mathrm{T}\mathbf{K}_j\bm{\Phi}_j
    =\frac{(-1)^{j+1}}{2}\rho_j\bm{\dot{\Xi}}^\mathrm{T}\mathbf{L}_j^\mathrm{T}\mathbf{K}_j\mathbf{L}_j\bm{\dot{\Xi}},
\end{equation}
where we used the definition of $\mathbf{K}_j$ (see \eqref{eq:app:kdefinition}), and the $(-1)^{j+1}$ accounts for the necessary change in sign of the $z-$integral in fluid $2$ to match the definition of $\mathbf{K}_j$ in equation~\eqref{eq:app:kdefinition}. The last term of the equation can be conveniently simplified by noting that $\mathbf{L}_j^\mathrm{T}\mathbf{K}_j\mathbf{L}_j=\mathbf{D}_j(\mathbf{K}_j^{-1})^\mathrm{T}\mathbf{K}_j\mathbf{L}_j=\mathbf{D}_j\mathbf{L}_j\equiv\mathbf{A}_j$. The exact form of matrix $\mathbf{A}_j$ is given in~\eqref{eq:app:Amatrix} of~\ref{app:MatrixInt}. Accordingly, the total kinetic energy of the fluids read
\begin{equation}
\label{eq:totKineticEn}
    \frac{T}{\Sigma} = \frac{1}{2}\bm{\dot{\Xi}}^\mathrm{T}\left(\rho_1\mathbf{A}_1-\rho_2\mathbf{A}_2\right)\bm{\dot{\Xi}}.
\end{equation}
To leading order of $\xi_c$ in the matrices $\mathbf{A}_j$, the equation above reduces to 
\begin{equation}
    \frac{T}{\Sigma} = \frac{1}{2}\sum_{a} \left(\frac{\rho_1}{k_a\tanh(k_a|h_1|)}+\frac{\rho_2}{k_a\tanh(k_a|h_2|)}\right)\dot{\xi}_a^2,
\end{equation}
and reveals the linear order interfacial dynamics. Going further, through the dependence of the matrices $\mathbf{A}_j$ on powers of $\xi_a$, cf.~\eqref{eq:app:Amatrix}, we see that the variational approach of the previous section allows us to include nonlinear terms in the Lagrangian formulation of the interfacial dynamics.

Additionally, the Lagrangian requires a potential energy $V$, which we obtain from the effective gravitational energy, $\rho_j g(t)z$, integrated over deformations $\xi(t,\bm{x})$ of the resting interface at $z=0$, and the surface energy term as a result of the non-vanishing interfacial tension $\sigma$. Hence, the energy reads
\begin{equation}
    \label{eq:potentialEnergy}
\begin{aligned}    
    \frac{V}{\Sigma} &= \frac{1}{2\Sigma}\iint\dd\Sigma~\left[\left(\rho_1-\rho_2\right)g(t)\xi^2+2\sigma\left(|\nabla\Gamma|-1\right)\right]\\
    &=\frac{1}{2\Sigma}\iint\dd\Sigma~\left[\left(\rho_1-\rho_2\right)g(t)\xi^2+\sigma\left((\nabla\xi)^2-\frac{1}{4}(\nabla\xi)^4+\cdots\right)\right].
\end{aligned}
\end{equation}
As presented here, the potential energy $V$ is defined so that it vanishes in the absence of interfacial waves $\xi(t,\bm{x})$. The last term in \eqref{eq:potentialEnergy} will introduce additional non-linear contributions to the Lagrangian, and similar to the derivations of the previous section, the quartic term in equation~\eqref{eq:potentialEnergy} may be expressed in the spectral decomposition~\eqref{eq:xidecomp} as
\begin{equation}
    \frac{1}{4\Sigma}\iint\dd\Sigma~(\nabla\xi)^4\equiv \frac{1}{4}\sum_{a,b,c,d}\mathcal{B}_{abcd}\xi_a\xi_b\xi_c\xi_d,
\end{equation}
where $\mathcal{B}_{abcd}$ are coefficients computed from products of $\nabla\chi_a$ and defined in~\eqref{eq:app:Bcoeff} of~\ref{app:MatrixInt}.

Provided with both the kinetic and potential energies, $T$ and $V$, we can then define the complete Lagrangian that determines the interfacial dynamics of the height modes $\xi_a$, as follows
\begin{multline}
\label{eq:nonlinLagrangia}
    \frac{L}{\Sigma}
    =\frac{1}{2}\sum_a\left(\frac{\rho_1}{k_aT_{1,a}}+\frac{\rho_2}{k_aT_{2,a}}\right)\left(\dot{\xi}_a^2-\omega_a^2(t)\xi_a^2\right)
    +\frac{1}{2}\sum_{a,b,c}\left(\rho_1\pow{\mathcal{A}}{1}_{cab}-\rho_2\pow{\mathcal{A}}{2}_{cab}\right)\xi_c\dot{\xi}_a\dot{\xi}_b\\
    +\frac{1}{4}\sum_{a,b,c,d}\left[\left(\rho_1\pow{\mathcal{A}}{1}_{cdab}
        +\rho_2\pow{\mathcal{A}}{2}_{cdab}\right)\dot{\xi}_a\dot{\xi}_b
        +\frac{\sigma}{2}\mathcal{B}_{abcd}\xi_a\xi_b\right]\xi_c\xi_d,
\end{multline}
where
\begin{equation}
\label{eq:dispersion}
    \omega_a^2(t)=\frac{(\rho_1-\rho_2)g(t)+\sigma k_a^2}{\rho_1\tanh(k_a|h_2|)+\rho_2\tanh(k_a|h_1|)}k_a\tanh(k_a|h_1|)\tanh(k_a|h_2|),
\end{equation}
and $T_{j,a}\equiv \tanh(k_a|h_j|)$. The time-dependent frequency $\omega_a^2(t)$ denotes the dispersion relation for an interfacial mode $\xi_a$ in linear dynamics obtained from leading-order terms of the Lagrangian (cf.~\cite{Kumar94}). By varying equation~\eqref{eq:nonlinLagrangia} with respect to $\xi_a$, one obtains the non-linear equations of motion for the interface in the presence of external forcing $F(t)$, which is contained within the effective gravity term $g(t)=g_0-F(t)$.

The coefficients $\mathcal{A}^{(j)}_{cab}$ and $\mathcal{A}^{(j)}_{cdab}$ in~\eqref{eq:nonlinLagrangia}  act as momentum conservation constrains. For a given primary $\xi_a$ mode with spatial wavenumber $k_a$, the values of the non-linear coefficients will determine which other modes will interact with the primary, similar to the conservation rules in vertices of Quantum Field Theory (QFT)~\cite{peskin2018introduction}. However, in our proposed system, finite-size effects imposed by the physical boundaries cannot be neglected, and hence the discrete spectrum of wavenumbers $k_a$ further restrains the available interactions. 

\section{Approximate cases and EFT simulators}\label{sec:simulators}

Sch{\"u}tzhold and Unruh showed in~\cite{schutzhold2002gravity} that the equations of motion for long-wavelength (shallow-water) surface waves propagating on a fluid flow are mathematically equivalent to the dynamics of a massless Klein-Gordon field on a curved spacetime. We consider now several limiting cases of the equations presented in \Sref{sec::interfacial dynamics}, wherein one finds similar mathematical equivalences. Throughout this section, unless otherwise stated, we assume both fluids have the same depth, the considered modes have long wavelengths and capillary effects are negligible, i.e., $h_2=|h_1|\equiv h_0$, and their wavenumbers $k_a$ satisfy $k_a h_0 \ll 1$ and $\sigma k_a^2\ll (\rho_1-\rho_2)g_0$.

\subsection{Analogue massless scalar field in flat spacetime}
Consider first our setup in \Sref{sec::interfacial dynamics} in the absence of external forcing $F(t)$. Under the above conditions, the dispersion relation~\eqref{eq:dispersion} reads
\begin{equation}
    \omega_a^2 = (A g_0 h_0) k_a^2\equiv c^2 k_a^2,
\end{equation}
where $A\equiv \tfrac{\rho_1-\rho_2}{\rho_1+\rho_2}$ is the Atwood number and $c$ is the effective interfacial wave speed of propagation; a linear dispersion relation as in standard QFT. Further, the equations of motion for the modes $\xi_a$ to leading order are
\begin{equation}
\label{eq:analogueMinkowski}
    \ddot{\xi}_a+c^2 k_a^2\xi_a=0 \Leftrightarrow \frac{1}{c^2}\partial^2_t\xi-\nabla^2\xi=0,
\end{equation}
in whose derivation we used decomposition~\eqref{eq:xidecomp}. This is readily recognisable as the equation governing a massless Klein-Gordon field in a $(2+1)-$dimensional Minkowski spacetime with an effective speed of light $c=\sqrt{A g_0 h_0}$. In the limit $\rho_1\gg\rho_2$, the Atwood number $A$ tends to unity, and the resulting equation represents the effective field theory for surface modes in an inviscid fluid. This correspondence is of particular relevance for creating gravity simulators of quantum processes that can take place on flat spacetime, such as the Unruh effect~\cite{PhysRevD.14.870}, as proposed in~\cite{BunneyCircular, bunney2023third} on the interface of superfluid helium.

\subsection{A simulator for inflationary cosmology}
Allowing again an arbitrary external forcing $F(t)\neq0$, we promote the velocity potential $\phi_j$ to the main field and remark that at the interface, we have $\left.\phi_1\right|_{z=\xi}=-\left.\phi_2\right|_{z=\xi}=\phi_0$. Linearising \eqref{eq:interfacialDyn} in the limit of negligible capillary effects, we find
\begin{equation}
    (\rho_1+\rho_2)\frac{\partial}{\partial t}\phi_0+(\rho_1-\rho_2)g(t)\xi=0.
\end{equation} Applying to this the decompositions \eqref{eq:phidecomp} and \eqref{eq:xidecomp} and dividing by $(\rho_1+\rho_2)$, one recovers
\begin{equation}\label{eq:linearised dynamics}
\sum_a\left(\dot{\phi}_{0,a}(t)\psi_{0,a}(\bm{x})+Ag(t)\xi_a(t)\chi_a(\bm{x})\right)=0.
\end{equation} We note that the spatial eigenfunctions $\psi_{j,a}$ can be further decomposed as $\psi_{j,a}(\bm{x},z)=\chi_a(\bm{x})f_{j,a}(z)$ where $f_{j,a}(\xi)= 1$ to leading order at the interface. Hence, \eqref{eq:linearised dynamics} becomes
\begin{align}
    \sum_a\left(\dot{\phi}_{0,a}+Ag(t)\xi_a(t)\right)\chi_a(\bm{x})&=0,\\
    \implies \dot{\phi}_{0,a}+Ag(t)\xi_a(t)&=0.\label{eq:reduced equation}
\end{align}
Differentiating \eqref{eq:reduced equation} with respect to time yields
\begin{equation}
    \ddot{\phi}_{0,a}(t)+A\dot{g}(t)\xi_a(t)+Ag(t)\dot{\xi}_a(t)=0.
\end{equation} As we consider $\phi_0$ the main quantity of interest, we want to eliminate any instances of $\xi_a$. First, we recall that $\dot{\xi}(t)=k_a\tanh(k_ah_0)\phi_{0,a}=(h_0k_a^2)\phi_{0,a}$ in the long-wavelength limit. Second, we rearrange \eqref{eq:reduced equation} as $\xi_a=-\tfrac{1}{Ag(t)}\dot{\phi}_{0,a}$. Accounting for these, the equation of motion for our scalar field is
\begin{equation}
    \ddot{\phi}_{0,a}-\frac{\dot{g}(t)}{g(t)}\dot{\phi}_{0,a}+Ah_0k_a^2\phi_{0,a}=0.
\end{equation} If we make the following identifications $\omega_a^2(t)=(Ag(t)h_0)k_a^2$ and $a^{-2}\equiv Ag(t)h_0$ and relabel the mode number from $a$ to $d$ to avoid confusion this equation of motion can be rewritten in two ways,
\begin{subequations}
    \begin{align}
      \ddot{\phi}_{0,d}-\frac{\dot{g}(t)}{g(t)}\dot{\phi}_{0,d}+\omega_d^2(t)\phi_{0,d}&=0,\\
      \ddot{\phi}_{0,d}+2\frac{\dot{a}(t)}{a(t)}\dot{\phi}_{0,d}+\frac{k^2_d}{a^2(t)}\phi_{0,d}&=0.\label{eq:FLRW scalar field}
    \end{align}
\end{subequations}

Consider for a moment the general case of a massive scalar field in a $(2+1)-$dimensional Friedmann–Lemaître– Robertson–Walker (FLRW) spacetime~\cite{birrell_davies_1982,hawking_ellis_1973}, described by the following metric and Lagrangian density,
\begin{subequations}
    \begin{align}
    \mathrm{d}s^2&=g_{\mu\nu}\mathrm{d}x^\mu\mathrm{d}x^\nu=-\mathrm{d}t^2+a^2(t)\mathrm{d}\bm{\mathsf{x}}^2,\\
    \mathcal{L}&=-\frac{1}{2}\left(\nabla_\mu\phi\nabla^\mu\phi+m^2\phi^2\right)\sqrt{-g},\label{eq:Lagrangian density}
\end{align} where $g=\det{g_{\mu\nu}}.$
\end{subequations} The Euler-Lagrange equations of \eqref{eq:Lagrangian density} are 
\begin{equation}
    \Box\phi-m^2\phi=0,
\end{equation} which for the metric above reduces to
\begin{equation}
    \ddot{\phi}+2\frac{\dot{a}}{a}\dot{\phi}-\frac{1}{a^2}\nabla^2\phi+m^2\phi=0.\label{eq:FLRW general}
\end{equation} 
Note the factor of two in the term $2\tfrac{\dot{a}}{a}\dot{\phi}$ originates from the fact we are in a $(2+1)-$dimensional spacetime. 

Comparing \eqref{eq:FLRW scalar field} with \eqref{eq:FLRW general} allows us to interpret \eqref{eq:FLRW scalar field} in the following way: a massless, minimally coupled scalar field in a $(2+1)-$dimensional FLRW spacetime spectrally decomposed as $\phi(t,\mathsf{x})=\sum_d \phi_d(t)\chi_d(\mathsf{x})$, with $(\nabla^2+k_d^2)\chi_d=0$. The direct equivalence between the analogue system and an FLRW spacetime, however, relies on the long-wavelength limit. Had we considered arbitrary wavelengths, $a(t)$ would instead have been replaced by $a_k(t)$, where the scale factor would then depend on the spatial wavenumber $k_d$, characteristic of a rainbow universe~\cite{Fifer2019AnalogField,Weinfurtner_2009}. As in the case of~\cite{Fifer2019AnalogField}, by appropriately choosing the external forcing $F(t)$, and hence modulating the effective scale factor $a(t)$, one can use this correspondence to investigate the evolution of the analogue scalar field in inflationary scenarios.

Whilst the equations of motion for $\phi_{0,a}$ in this limiting case lead to simulating an FLRW spacetime, we bring to the attention of the reader the equations of motion for $\xi_a$. Note that equation~\eqref{eq:reduced equation} provides an interpretation of the field $\xi$ as the conjugate momentum of the effective field $\phi_0$ at the interface. Consider again the relation $\dot{\xi}_a=k_a\tanh(k_ah_0)\phi_{0,a}$, which simplifies to $\dot{\xi}_d=(h_0k_a^2)\phi_{0,a}$ under the present assumptions. Differentiating this yields
\begin{equation}
    \ddot{\xi}_a=h_0k_a^2\dot{\phi}_{0,a}.
\end{equation} We can rewrite this wholly in terms of $\xi_a$ by using \eqref{eq:reduced equation},
\begin{equation}\label{eq:xi eom}
    \ddot{\xi}_a+(Ag(t)h_0)k_a^2\xi_a=0.
\end{equation}
It is worth contrasting~\eqref{eq:analogueMinkowski} and~\eqref{eq:xi eom} and note that the latter is the equation of motion of the conjugate momentum of the analogue relativistic field $\phi_0$ in an effective FLRW spacetime with the speed of light set to unity. Whereas the latter describes the dynamics of the analogue height field $\xi$ in a Minkowski metric with an effective speed of light given by the fluid's wave speed of propagation. 

\subsection{Nonlinear EFT simulator}
We now specialise in the case of a harmonic external forcing $F(t)=F_0\cos(\Omega t)$\footnote{Here, we implicitly require that $F_0<g_0$, i.e., the amplitude of the external oscillating acceleration must not overcome Earth's gravitational acceleration $g_0$.}, for which the equation of motion of the modes $\xi_a$ \eqref{eq:xi eom} reads
\begin{subequations}
\label{eq:analogueMR}
\begin{align}
    \ddot{\xi}_a+k_a^2c^2\left(1-\frac{F_0}{g_0}\cos(\Omega t)\right)\xi_a&=0\label{eq:mathieu}\\
    \Rightarrow\frac{1}{\tilde{c}^{2}}\ddot{\xi}_a+\left(k_a^2+\lambda^2 k_a^2 F_0 \sin^2(\omega_0 t)\right)\xi_a&=0,\label{eq:analogueRehea}    
\end{align}
\end{subequations}
where we identified $\Omega\equiv 2\omega_0$, $\tilde{c}^2\equiv A(g_0-F_0)h_0$ and $\lambda^2\equiv 2/(g_0-F_0)$. In line with the previous discussion, we note that the equation above may be recast into that of a massless Klein-Gordon field propagating on a $(2+1)-$dimensional Minkowski spacetime, with an effective speed of light $\tilde{c}$. The analogue field $\xi$ is subject to a derivative coupling with a spatially uniform mechanical forcing $\Phi^2(t)$ through an interaction potential of the form $-\tfrac{1}{2}\lambda^2\Phi^2\left(\nabla\xi\right)^2$. Upon identifying $\Phi(t)\equiv\sqrt{F_0}\sin(\omega_0 t)$, one can readily recover equation~\eqref{eq:analogueRehea}.

Equation~\eqref{eq:mathieu} is known as the Mathieu equation~\cite{Kovacic2018MathieusFeatures,Magnus2013HillsEquation}, and its set of exponentially unstable solutions in time appears in a wide range of fields in physics, from engineering~\cite{Gazzola2015BriefBridges,Biran2014ChapterWaves} and fluid dynamics~\cite{Faraday1831XVII.Surfaces,Kumar94} to models of the Early Universe~\cite{Kofman94,Shtanov:1994ce}. In general, systems displaying these solutions are said to be undergoing parametric amplification or resonance, and in the context of fluid mechanics, the resonant modes are referred to as Faraday instabilities~\cite{Faraday1831XVII.Surfaces}. Building on our previous discussion, the Mathieu equation allows us to draw the comparison between our effective field theory of interfacial waves and models for the thermalisation of the Early Universe after inflation. 

In chaotic inflation models~\cite{Mukhanov2005PhysicalCosmology}, a scalar field $\varphi$, referred to as the inflaton, is under the action of any slow-roll potential $V(\varphi)$, and sources inflation on an FLRW spacetime with scale factor $a(t)$ and Hubble parameter $H\equiv \dot{a}/a$. During this process, the field undergoes a classical background evolution according to the inflationary expansion. By the end of inflation, the inflaton leaves the potential dominated region and oscillates around the global minimum of $V(\phi)$, starting reheating. We consider a massless bosonic field $\psi$ coupled to the inflaton $\varphi$ via the interaction potential $-\tfrac{1}{2}\lambda\varphi^2\psi^2$. 

After inflation, the field $\varphi$ is assumed to be oscillating around the minimum of a quadratic potential $\tfrac{1}{2}m_{\varphi}^2\varphi^2$, i.e., $\varphi(t)\approx \Phi(t) \sin(m_{\varphi}t)$, where $\Phi(t)$ is a nearly constant amplitude. One can then consider the oscillation energy of the inflaton transferring to the matter field $\psi$ through their coupling, resulting in the following equation of motion for $\psi_k$ modes~\cite{Kofman94,Baumann2015TheInflation}
\begin{equation}
    \ddot{\psi}_k+3H\dot{\psi}_k +\left(\frac{k^2}{a^2(t)}+\lambda^2\varphi^2(t)\right)\psi_k=0.\label{eq:reheatEOM1}
\end{equation}
In these conditions, the spacetime may be treated as a nearly Minkowski metric with $a(t)\sim 1$ and $ \dot{a}(t)\sim 0$~\cite{Mukhanov2005PhysicalCosmology}. Thus, we can neglect the Hubble-friction term in the equation above, which reduces to 
\begin{subequations}
    \begin{align}
        \ddot{\psi}_k+\left(k^2+\lambda^2\Phi^2\sin^2(m_{\varphi}t)\right)\psi_k&=0\label{eq:reheatDyn}\\
        \Rightarrow \ddot{\psi}_k+\left(k^2+\frac{1}{2}\lambda^2\Phi^2-\frac{1}{2}\lambda^2\Phi^2\cos(2m_{\varphi}t)\right)\psi_k&=0,\label{eq:reheatMathieu}
    \end{align}
\end{subequations}
resulting in a Mathieu-type equation as in~\eqref{eq:analogueMR}.

For certain bands of $k$-values, the modes $\psi_k$ experience instabilities in the form of
\begin{equation}
    \psi_k (t)\propto \exp(\mu_k m_{\varphi} t),
\end{equation}
where $\mu_k$ are known as Floquet coefficients. For large values of $\lambda^2\Phi^2$ in equation \eqref{eq:reheatMathieu}, a broadband of $k$-modes enters resonance, which characterises the so-called preheating~\cite{Kofman94}, in the initial stages of reheating. The unstable behaviour results in the explosive production of $\psi-$particles, whose occupation number density $n_k$ grows exponentially with the instability rate $\mu_k$, i.e., $n_k\propto \exp(2\mu_k m_\varphi t)$. The outcome of preheating is a far-from-equilibrium state, which is subject to elementary particle decays, leading to subsequent stages of reheating and, finally, the required thermalisation of the Early Universe~\cite{Mukhanov2005PhysicalCosmology}. 

We now turn to the equivalence between equations~\eqref{eq:analogueMR} and~\eqref{eq:reheatMathieu} and consider scenarios in which Faraday instabilities can be parametrically excited on the fluid interface by the external forcing $F(t)$. Within the long-wavelength limit and neglecting capillary effects, the Lagrangian~\eqref{eq:nonlinLagrangia} to second-order in small amplitudes $\xi_a$ reduces to
\begin{equation}
\label{eq:nonlinLagrangia2}
    \frac{L}{\Sigma}
    =\frac{\rho_1+\rho_2}{2h_0}\sum_a\frac{1}{k_a^2}\left(\dot{\xi}_a^2-\omega_a^2(t)\xi_a^2\right)
    +\frac{\rho_1-\rho_2}{2}\sum_{a,b,c}\pow{\mathcal{A}}{0}_{cab}\xi_c\dot{\xi}_a\dot{\xi}_b
    +\frac{\rho_1+\rho_2}{4}\sum_{a,b,c,d}\pow{\mathcal{A}}{0}_{cdab}\dot{\xi}_a\dot{\xi}_b\xi_c\xi_d,
\end{equation}
with $\omega_a^2(t)=(Ag(t)h_0)k_a^2$. By evaluating the Euler-Lagrange equations for a mode $\xi_a$, one obtains its nonlinear equation of motion, which recovers~\eqref{eq:mathieu} to leading order. The validity of the resulting equation relies on the modes satisfying both the long-wavelength and negligible capillarity conditions, which may strongly restrict the wavenumbers that can enter the approximate dynamics. Regardless, one can realise a scenario where a low-wavenumber interfacial mode undergoes parametric amplification and evolves dominantly with respect to the others. As its amplitude exponentially grows over time, other modes may be excited through the cubic and quartic interactions in equation~\eqref{eq:nonlinLagrangia2}. 

In this picture, beyond the correspondence of the interfacial height $\xi$ with a Klein-Gordon field on flat spacetime, the emerging EFT given by Lagrangian~\eqref{eq:nonlinLagrangia2} can be used to investigate the onset of interactions in a simulator for preheating and the early stages of the thermalisation of the Early Universe~\cite{barroso2022primary}. One must keep in mind, however, that the geometry of the fluid cell determines the values of $k_a$ that are available through the boundary conditions~\eqref{eq:nopen} on the analogue field $\xi$. That amounts to saying that there is a reduced density of states $\xi_a$ given by the discrete set of $k_a$ modes that fit on the confined interface. 

To illustrate the implications of discretisation, we consider a fluid cell with a square cross-section of sides $L$, i.e., $(x,y)\in \left[0,L\right]\times\left[0,L\right]$ and $\Sigma=L^2$. In this geometry, the spatial eigenfunctions $\chi_{mn}(\bm{x})$ satisfying~\eqref{eq:horizontalHelm} read
\begin{equation}
\label{eq:eifunctions square}
    \chi_{mn}(x,y) = 2\cos\left(\frac{\pi m x}{L}\right)\cos\left(\frac{\pi n y}{L}\right),
\end{equation}
where $m,~n\in\mathbb{N}\cup\{0\}$ such that $m^2+n^2=N>0$. The corresponding eigenvalues of $\chi_{mn}(\bm{x})$ are $k_{mn}=\tfrac{\sqrt{N}\pi}{L}$. The pair of non-negative integers $(m,n)$ uniquely defines interfacial modes $\xi_{mn}(t)$ with a degenerate spectrum and wavevector $\bm{k}_{mn}=\tfrac{\pi}{L}(m\unvec{x}+n\unvec{y})$. This degeneracy arises from the invariance of $\sqrt{N}$ under the action $(m,n)\mapsto(n,m)$; equivalently, this is to be anticipated from the square geometry of the system which enjoys a discrete symmetry under rotations of $\pi/2$~\footnote{If, however, the geometry is taken to be $(x,y)\in[0,L_x]\times[0,L_y]$ with $L_x\neq L_y$, then this symmetry is broken and the spectrum is non-degenerate with unique pair labels $(m,n)$.}. With respect to this geometry and to satisfy our assumptions, long-wavelength modes must then satisfy $\sqrt{N}\ll \tfrac{L}{\pi h_0}$, which is guaranteed by requiring a configuration with large aspect ratio $L/h_0$.  

By computing the coefficients $\pow{\mathcal{A}}{0}_{lmn,l' m' n'}$ and $\pow{\mathcal{A}}{0}_{rlmn,r' l' m' n'}$ in~\ref{app:MatrixInt} using $\chi_{m,m'}$ of the form~\eqref{eq:eifunctions square}, we find that
\begin{subequations}
\label{eq:square conservation}
\begin{align}
    \pow{\mathcal{A}}{0}_{lmn,l' m' n'} &\propto \delta_{l,|m\pm n|}\delta_{l',|m'\pm n'|}, \\
    \pow{\mathcal{A}}{0}_{rlmn,r' l' m' n'} &\propto \delta_{|m\pm r|,|n\pm l|}\delta_{|m'\pm r'|,|n'\pm l'|},
\end{align}    
\end{subequations}
where primes indicate the mode-numbers $(r,l,m,n)$ in the $y-$direction, and $\delta_{ab}$ is the Kronecker delta. Equations~\eqref{eq:square conservation} correspond to conservation rules on the Lagrangian~\eqref{eq:nonlinLagrangia2}, as they effectively restrict the contributions to cubic and quartic terms to modes satisfying the delta relations above. For instance, the cubic constrain requires that a mode with wavenumber $k_{l,l'}$ can only decay into wavenumbers $k_{m,m'}$ and $k_{n,n'}$, if $l=|m\pm n|$ and $l'=|m'\pm n'|$. In the context of our interacting EFT, the deltas in equations~\eqref{eq:square conservation} are equivalent to vertex conservation rules in standard QFT. 

Along these lines, the nonlinear evolution of an interfacial mode $\xi_{mm'}(t)$ is given by
\begin{equation}
    \ddot{\xi}_{mm'}+\omega_{mm'}^2(t)\xi_{mm'}+ \frac{2h_0k_{mm'}^2}{\rho_1+\rho_2}\left[\frac{d}{dt}\left(\frac{\partial \pow{L}{1}}{\partial \dot{\xi}_{mm'}}\right)-\frac{\partial \pow{L}{1}}{\partial \xi_{mm'}}\right]=0,
\end{equation}
where $\pow{L}{1}$ correspond to cubic and higher-order terms in the Lagrangian~\eqref{eq:nonlinLagrangia2}. With the appropriate selection of a dominant mode, one can simplify this equation to include only terms resulting from quartic interactions in the Lagrangian and hence simulate $\phi^4-$like models for the particle decay of reheating, as proposed in~\cite{barroso2022primary}. Generally, the formalism presented here offers the possibility of investigating dynamical features of interacting field theories with the emerging description of nonlinearities on fluid interfaces. 

\section{Conclusion}

We presented in this Proceeding a detailed derivation of an effective field theory for the nonlinear dynamics of interfacial waves between two fluids based on prior work on fluid dynamics~\cite{Miles1976NonlinearBasins,Miles1984NonlinearResonance}. By focusing on specific approximate cases, we reviewed previous linear simulators for flat and inflationary spacetimes as in~\cite{bunney2023third} and~\cite{Fifer2019AnalogField}, and built-up a framework that can be used to investigate interacting field theories. In the latter case, we exemplified this by simulating the nonlinear onset and subsequent particle decay of reheating in inflationary cosmology, in line with~\cite{barroso2022primary}.

We note that the assumptions required for some of the derivations in~\Sref{sec::interfacial dynamics} and~\Sref{sec:simulators} may not be applicable in all experimental realisations. However, the formalism can be accordingly modified to improve the description of the underlying physical system used as a simulator and, in turn, reveal novel aspects of the emerging EFT. For instance, in most classical fluids, the contribution of linear damping to the equations of motion of interfacial waves is not generally negligible; however, the Lagrangian theory can then be modified by the inclusion of a Rayleigh dissipation function~\cite{barroso2022primary,Miles1984NonlinearResonance}.

When designing gravity simulators, one must be aware that mathematical equivalences may not be achievable from the underlying theory or realisable in experiments within their regimes of validity. Nonetheless, analogue systems offer the possibility of examining the universality and robustness of the theoretical descriptions that they aim to simulate. Notable examples include measurements of analogue Hawking radiation~\cite{Hawking1975ParticleHoles} around various effective horizons~\cite{Belgiorno2010HawkingFilaments,Philbin2008Fiber-opticalHorizon,Weinfurtner2011MeasurementSystem,Steinhauer2016ObservationHole}, and analogue black hole phenomena in dispersive draining vortex flows~\cite{Torres2017RotationalFlow,Torres20QNM}. Building on the firm foundations laid by these previous simulators, recent developments are motivating and driving the extension of the underlying quantum field theories~\cite{Fifer2019AnalogField,barroso2022primary,BunneyCircular,bunney2023third}, fostering a symbiosis between experiment and theory within analogue gravity.

\ack
We thank the organisers of the ``Avenues of Quantum Field Theory in Curved Spacetime" workshop for the exciting, multifaceted showcase of the recent developments in QTFCS, which inspired an extension of the work presented there. 
The authors are grateful for fruitful discussions with August Geelmuyden and all those involved in the projects that led to this work. 
SW acknowledges support provided by the Leverhulme Research Leadership Award (RL-2019- 020),
the Royal Society University Research Fellowship
(UF120112, RF\textbackslash ERE\textbackslash 210198) and the Royal Society Enhancements Awards and Grants
(RGF\textbackslash EA\textbackslash 180286, RGF\textbackslash EA\textbackslash 181015), and partial
support by the Science and Technology Facilities Council (Theory Consolidated Grant ST/P000703/1), the Science and Technology Facilities Council on Quantum Simulators for Fundamental Physics (ST/T006900/1) as part
of the UKRI Quantum Technologies for Fundamental
Physics programme.

\section*{References}
\bibliographystyle{ieeetr} 
\bibliography{refs} 

\appendix 
\section{Variation of the kinematic boundary action}
\label{app:variationAction}
We consider the action defined in equation~\eqref{eq:actionDirichlet},
\begin{equation}
\begin{aligned}    
    I_j &= \frac{(-1)^{j+1}}{2\Sigma}\iiint_{V_j} \left(\nabla\phi_j\right)^2 \dd\Sigma~\dd z
    -\frac{1}{\Sigma}\iint \frac{1}{|\nabla\Gamma|}\frac{\partial\xi}{\partial t}\left.\phi_j\right|_{z=\xi} \dd\Sigma\\
    &=\frac{(-1)^{j+1}}{2\Sigma}\iint\dd\Sigma(-1)^{j+1}\int_{h_j}^{\xi}\dd z~\left(\nabla\phi_j\right)^2-\frac{1}{\Sigma}\iint \frac{1}{|\nabla\Gamma|}\frac{\partial\xi}{\partial t}\left.\phi_j\right|_{z=\xi} \dd\Sigma.
\end{aligned}
\end{equation}
The variation of the action with respect to $\phi_j$ is then given by
\begin{equation}
\begin{aligned}    
    \delta I_j &= \frac{(-1)^{j+1}}{\Sigma}\iiint_{V_j}\dd\Sigma\dd z ~\nabla\phi_j\cdot\nabla(\delta\phi_j)-\frac{1}{\Sigma}\iint\dd\Sigma~\frac{1}{|\nabla\Gamma|}\frac{\partial\xi}{\partial t}\left.\delta\phi_j\right|_{z=\xi}\\
    &=\frac{(-1)^{j+1}}{\Sigma}\iiint_{V_j}\dd\Sigma\dd z~\left[\nabla\cdot\left(\delta\phi_j\nabla\phi_j\right)-\delta\phi_j\nabla^2\phi_j\right]
    -\frac{1}{\Sigma}\iint\dd\Sigma~\frac{1}{|\nabla\Gamma|}\frac{\partial\xi}{\partial t}\left.\delta\phi_j\right|_{z=\xi}\\
    &=\frac{(-1)^{j}}{\Sigma}\iiint_{V_j}\dd\Sigma\dd z~\left(\nabla^2\phi_j\right)\delta\phi_j
    +\frac{(-1)^{j+1}}{\Sigma}\iint_{\partial V_j}\dd S_{\unvec{n}}~\left[\left(\unvec{n}\cdot\nabla\phi_j\right)\delta\phi_j\right]_{\partial V_j}\\
    &\hspace{250pt}-\frac{1}{\Sigma}\iint\dd\Sigma~\frac{1}{|\nabla\Gamma|}\frac{\partial\xi}{\partial t}\left.\delta\phi_j\right|_{z=\xi}.
\end{aligned}
\end{equation}
The vector $\unvec{n}$ denotes all normal unit vectors of the boundaries $\partial V_j$ of volume $V_j$ of each fluid with $S_{\unvec{n}}$ their respective area element. We distinguish the boundaries into rigid ones, $\overline{\partial V_j}$, and the moving interface at $\Gamma=0$. For the latter, the outwardly directed normal $\unvec{n}_{\Gamma,j}$ of each fluid can be written as $\unvec{n}_{\Gamma,j}=(-1)^{j+1}\unvec{n}_\Gamma=(-1)^{j+1}\nabla\Gamma/|\nabla\Gamma|$, i.e., the interface's normal points upwards in fluid $1$ and downwards in $2$. Hence, the variation of the action results in
\begin{multline}
    \delta I_j =\frac{(-1)^{j}}{\Sigma}\iiint_{V_j}\dd\Sigma\dd z~\left(\nabla^2\phi_j\right)\delta\phi_j
    +\frac{(-1)^{j+1}}{\Sigma}\iint_{\overline{\partial V_j}}\dd S_{\unvec{n}}~\left[\left(\unvec{n}\cdot\nabla\phi_j\right)\delta\phi_j\right]_{\overline{\partial V_j}}\\
    +\frac{1}{\Sigma}\iint\dd\Sigma\frac{1}{|\nabla\Gamma|}\left[\nabla\Gamma\cdot\nabla\phi_j-\frac{\partial\xi}{\partial t}\right]_{z=\xi}\delta\phi_j|_{z=\xi}.
\end{multline}
Hence, by imposing $\delta I_j=0$, we see that the integrals in the equation above respectively recover Laplace's equation~\eqref{eq:Laplace} for $\phi_j$, the no-penetration conditions~\eqref{eq:nopen} at rigid boundaries, and the kinematic condition~\eqref{eq:kinematic} at the interface.

\section{$\bm{\mathrm{K}}_j$, $\bm{\mathrm{D}}_j$, $\bm{\mathrm{L}}_j$ and $\mathbf{A}_j$ matrices, their integral coefficients and useful relations}
\label{app:MatrixInt}

By comparing equation~\eqref{eq:action} with~\eqref{eq:actionDirichlet}, we can obtain the form of the $\bm{\mathrm{K}}_j$ and $\bm{\mathrm{D}}_j$ matrices. The matrix coefficients of $\bm{\mathrm{D}}_j$, denoted $\pow{\mathdutchcal{d}}{j}_{ab}$, are defined by
\begin{equation}
\label{eq:app:ddefinition}
    \pow{\mathcal{d}}{j}_{ab}=\frac{1}{\Sigma}\iint \dd\Sigma~ \frac{\chi_a\chi_b}{\sqrt{1+|\nabla\xi|^2}} \frac{\cosh[k_b(\xi-h_j)]}{\cosh(k_bh_j)}.
\end{equation}
This definition indicates that $\bm{\mathrm{D}}_j$ is an asymmetric, dimensionless, square matrix.
Under the assumption that $|\xi|\ll |h_j|$, we may right the hyperbolic cosines above as
\begin{equation}
    \frac{\cosh[k_b(\xi-h_j)]}{\cosh(k_bh_j)}= 1 + (-1)^{j+1}T_{j,b} k_b\xi +\frac{1}{2}k_b^2\xi^2 + \cdots,
\end{equation}
with $T_{j,a}\equiv \tanh(k_a|h_j|)$ and we used $h_j=(-1)^j |h_j|$. We must also account for the square root in the denominator of equation~\eqref{eq:app:ddefinition}, which for small slopes of the interface, i.e., $|\nabla\xi|\ll 1$, reads
\begin{equation}
    \frac{1}{\sqrt{1+|\nabla\xi|^2}}=1-\frac{1}{2}|\nabla\xi|^2+\cdots=1-\frac{1}{2}\sum_{c,d}\nabla\chi_c\cdot\nabla\chi_d \xi_c\xi_d+\cdots.
\end{equation}
From the equation above and~\eqref{eq:xidecomp}, we see that
\begin{equation}
\label{eq:app:dmatrix}
    \pow{\mathdutchcal{d}}{j}_{ab}=\delta_{ab}+(-1)^{j+1}k_bT_{j,b}\sum_c \mathbb{C}_{cab}\xi_c+\frac{1}{2}\sum_{c,d}\left(k_b^2\mathbb{C}_{cdab}-\mathbb{D}_{abcd}\right)\xi_c\xi_d+\cdots,
\end{equation}
with 
\begin{equation}
    \mathbb{C}_{cab} = \frac{1}{\Sigma}\iint \dd \Sigma~ \chi_c\chi_a\chi_b\text{, } \mathbb{C}_{cdab} = \frac{1}{\Sigma}\iint \dd \Sigma~ \chi_c\chi_d\chi_a\chi_b\text{ and }\mathbb{D}_{abcd} = \frac{1}{\Sigma}\iint \dd \Sigma~ \chi_a\chi_b\nabla\chi_c\cdot\nabla\chi_d~.
\end{equation}

Similarly, the matrix coefficients of $\bm{\mathrm{K}}_j$, denoted $\pow{\mathdutchcal{k}}{j}_{ab}$, are defined by
\begin{equation}
\label{eq:app:kdefinition}
    \pow{\mathdutchcal{k}}{j}_{ab}=\frac{(-1)^{j+1}}{\Sigma}\iiint_{V_j} \dd\Sigma\dd z~ \nabla\psi_a\cdot\nabla\psi_b=\frac{1}{\Sigma}\int_{h_j}^{\xi}\dd z\iint \dd\Sigma~ \nabla\psi_a\cdot\nabla\psi_b.
\end{equation}
It is worth noting that in fluid $2$ the integration in $z$ should span from $\xi$ to $h_2$, hence requiring a negative sign, which cancels off the sign in equation~\eqref{eq:actionDirichlet}.
$\bm{\mathrm{K}}_j$ is also a square matrix, but symmetric in its entries and with the dimension of inverse length. By using the definitions in the main text, the equation above expands to
\begin{equation}
\begin{aligned}
    \pow{\mathcal{k}}{j}_{ab}=\frac{\sech(k_a|h_j|)}{\Sigma\cosh(k_b|h_j|)}\int_{h_j}^{\xi}\dd z\iint \dd\Sigma~ \Bigl\{&\nabla\chi_a\cdot\nabla\chi_b \cosh[k_a(z-h_j)]\cosh[k_b(z-h_j)]\\
    &+k_ak_b\chi_a\chi_b \sinh[k_a(z-h_j)]\sinh[k_b(z-h_j)]\Bigl\},
\end{aligned}
\end{equation}
and upon integration of the $z-$components, reduces to
\begin{equation}
    \pow{\mathcal{k}}{j}_{ab}=\frac{1}{2\Sigma}\iint \dd\Sigma~ \Bigl\{\pow{\mathcal{K}}{j,+}_{ab}+ \pow{\mathcal{K}}{j,-}_{ab}\Bigl\},
\end{equation}
with
\begin{equation}
    \pow{\mathcal{K}}{j,\pm}_{ab}= \frac{\sech(k_a|h_j|)}{\cosh(k_b|h_j|)}\frac{\sinh[(k_a\pm k_b)(\xi-h_j)]}{k_a\pm k_b}\Bigl\{\nabla\chi_a\cdot\nabla\chi_b\pm k_ak_b\chi_a\chi_b\Bigl\}.
\end{equation}
And again, under the assumption that $|\xi|\ll |h_j|$, we find that
\begin{equation}
    \begin{aligned}
    \frac{\sinh[(k_a\pm k_b)(\xi-h_j)]}{k_a\pm k_b} 
    = (-1)^{j+1}&\frac{\sinh[(k_a\pm k_b)|h_j|]}{k_a\pm k_b}
    +\xi \cosh [(k_a\pm k_b)|h_j|]\\
    &+\frac{(-1)^{j+1}}{2} (k_a\pm k_b)\xi ^2 \sinh[(k_a\pm k_b)|h_j|]+\cdots,
    \end{aligned}
\end{equation}
which leads to 
\begin{equation}
    \begin{aligned}
    \pow{\mathcal{K}}{j,\pm}_{ab}
    = &(-1)^{j+1}\frac{T_{j,a}\pm T_{j,b}}{k_a\pm k_b}\Bigl\{\nabla\chi_a\cdot\nabla\chi_b\pm k_ak_b\chi_a\chi_b\Bigl\}\\
    &+(1\pm T_{j,a} T_{j,b})\sum_c \xi_c \Bigl\{\chi_c\nabla\chi_a\cdot\nabla\chi_b\pm k_ak_b\chi_c\chi_a\chi_b\Bigl\} \\
    &+\frac{(-1)^{j+1}}{2} (k_a\pm k_b) (T_{j,a}\pm T_{j,b})\sum_{c,d} \xi_c\xi_d \Bigl\{\chi_c\chi_d\nabla\chi_a\cdot\nabla\chi_b\pm k_ak_b\chi_c\chi_d\chi_a\chi_b\Bigl\} +\cdots.
    \end{aligned}
\end{equation}

By rearranging terms and defining new auxiliary integral coefficients, we find that the matrix coefficients $\pow{\mathcal{k}}{j}_{ab}$ up to second-order in $\xi$ simplify to 
\begin{equation}
\begin{aligned}
\label{eq:app:kmatrix}
    \pow{\mathcal{k}}{j}_{ab}=
    &(-1)^{j+1}k_a T_{j,a} \delta_{ab} 
    +\sum_c(\mathbb{D}_{cab}+k_aT_{j,a} k_b T_{j,b} \mathbb{C}_{cab})\xi_c\\
    &+\frac{(-1)^{j+1}}{2}\sum_{c,d} \Bigl[(k_aT_{j,a}+k_bT_{j,b})\mathbb{D}_{cdab}+(k_a^2k_bT_{j,b}+k_b^2k_aT_{j,a})\mathbb{C}_{cdab}\Bigl]\xi_c\xi_d
    +\cdots,
\end{aligned}
\end{equation}
with
\begin{equation}
    \mathbb{D}_{cab} = \frac{1}{\Sigma}\iint \dd \Sigma~ \chi_c\nabla\chi_a\cdot\nabla\chi_b\text{ and } \mathbb{D}_{cdab} = \frac{1}{\Sigma}\iint \dd \Sigma~ \chi_c\chi_d\nabla\chi_a\cdot\nabla\chi_b~.
\end{equation}
In deriving the equations above, one should note that the boundary condition on rigid boundaries~\eqref{eq:horizontalHelm} is required when setting boundary terms in the integrals to zero. Thus, if those conditions change, one should account for that when re-deriving the form of the matrix elements $\pow{\mathcal{k}}{j}_{ab}$. We also note that matrix $\bm{\mathrm{K}}_j$ in equation~\eqref{eq:app:kmatrix} agrees with the result from Miles~\cite{Miles1976NonlinearBasins} for the fluid $j=1$, however, matrix $\bm{\mathrm{D}}_j$ in equation~\eqref{eq:app:dmatrix} diverges from that of Miles by a term proportional to $\mathbb{D}_{abcd}$.

To obtain the $\mathbf{L}_j$ matrix coefficients, we must first compute the inverse of $\mathbf{K}_j$, which can be done by employing a matrix inversion identity (e.g., the Woodbury identity) or computing each term of the inverse perturbatively for a given small parameter. We opt for the latter and suppose that there exists a square matrix $\mathbf{F}_j$ such that $\mathbf{F}_j\mathbf{K}_j=\mathbf{I}$, where $\mathbf{I}$ is the identity matrix, and hence $\mathbf{F}_j\equiv(\mathbf{K}_j^{-1})^\mathrm{T}$. Similar to equation~\eqref{eq:app:kmatrix}, we may write the matrix coefficient of $\mathbf{F}_j$, denoted $\pow{\mathcal{f}}{j}_{ab}$, as a series expansion in powers of $\xi_a$, as follows,
\begin{equation}
    \pow{\mathcal{f}}{j}_{ab}= \pow{\mathcal{f}}{j,0}_{ab}
    +\sum_{c}\pow{\mathcal{f}}{j,1}_{abc}\xi_c
    +\sum_{c,d}\pow{\mathcal{f}}{j,2}_{abcd}\xi_c\xi_d+\cdots.
\end{equation}
By solving the equality $\sum_c\pow{\mathcal{f}}{j}_{ac}\pow{\mathcal{k}}{j}_{cb}=\delta_{ab}$ up to second order in $\xi_a$, we find
\begin{equation}
\begin{aligned}
\label{eq:app:fmatrix}
    \pow{\mathcal{f}}{j}_{ab}=
    &\frac{(-1)^{j+1}}{k_a T_{j,a}} \delta_{ab} 
    -\sum_c\left(\frac{\mathbb{D}_{cab}}{k_aT_{j,a} k_b T_{j,b}}+ \mathbb{C}_{cab}\right)\xi_c\\
    &+(-1)^{j}\frac{1}{2k_aT_{j,a}k_bT_{j,b}}\sum_{c,d} \Bigl[(k_aT_{j,a}+k_bT_{j,b})\mathbb{D}_{cdab}+(k_a^2k_bT_{j,b}+k_b^2k_aT_{j,a})\mathbb{C}_{cdab}\Bigl]\xi_c\xi_d\\
    &-(-1)^{j}\sum_{c,d,e}\frac{(\mathbb{D}_{cae}+k_aT_{j,a} k_e T_{j,e} \mathbb{C}_{cae})(\mathbb{D}_{deb}+k_eT_{j,e} k_b T_{j,b} \mathbb{C}_{deb})}{k_aT_{j,a}k_bT_{j,b}k_eT_{j,e}}\xi_c\xi_d
    +\cdots.
\end{aligned}
\end{equation}
Finally, the matrix elements $\pow{\mathcal{l}}{j}_{ab}$ can be found by computing $\pow{\mathcal{l}}{j}_{ab}=\sum_c \pow{\mathcal{f}}{j}_{ac}\pow{\mathcal{d}}{j}_{bc}$, which results in
\begin{equation}
\begin{aligned}
\label{eq:app:lmatrix}
    \pow{\mathcal{l}}{j}_{ab}=
    &\frac{(-1)^{j+1}}{k_a T_{j,a}} \delta_{ab} 
    -\sum_c\frac{\mathbb{D}_{cab}}{k_aT_{j,a} k_b T_{j,b}}\xi_c\\
    &+(-1)^{j+1}\sum_{c,d}\biggl[-\frac{k_b}{2T_{j,b}}\mathbb{C}_{cdab}-\frac{k_aT_{j,a}+k_bT_{j,b}}{2k_aT_{j,a}k_bT_{j,b}}\mathbb{D}_{cdab}-\frac{1}{2k_a T_{j,a}}\mathbb{D}_{abcd}\\
    &\qquad\qquad\qquad+\sum_{e}\frac{\mathbb{D}_{deb}}{k_aT_{j,a}k_bT_{j,b}k_eT_{j,e}}(\mathbb{D}_{cae}+k_aT_{j,a} k_e T_{j,e} \mathbb{C}_{cae})\biggl]\xi_c\xi_d
    +\cdots.
\end{aligned}
\end{equation}

The elements $\pow{\mathcal{A}}{j}_{ab}$ of matrix $\mathbf{A}_j$ may be computed in a similar way as $\mathbf{L}_j$ and $\mathbf{F}_j$ by defining it from the series expansion in powers of $\xi_a$, i.e.,
\begin{equation}
\label{eq:app:Amatrix}
    \pow{\mathcal{A}}{j}_{ab}= (-1)^{j+1}\pow{\mathcal{A}}{j,0}_{ab}
    +\sum_{c}\pow{\mathcal{A}}{j}_{cab}\xi_c
    +\frac{1}{2}(-1)^{j+1}\sum_{c,d}\pow{\mathcal{A}}{j}_{cdab}\xi_c\xi_d+\cdots,
\end{equation}
with
\begin{subequations}
\label{eq:app:AmatrixCoeffs}
\begin{align}
    \pow{\mathcal{A}}{j,0}_{ab}&=\frac{1}{k_aT_{j,a}}\delta_{ab}~,\\
    \pow{\mathcal{A}}{j}_{cab}&=\mathbb{C}_{cab}-\frac{\mathbb{D}_{cab}}{k_aT_{j,a}k_bT_{j,b}}~,\\
    \pow{\mathcal{A}}{j}_{cdab}&=-\frac{k_aT_{j,a}+k_bT_{j,b}}{k_aT_{j,a}k_bT_{j,b}}\left(\mathbb{D}_{abcd}+\mathbb{D}_{cdab}\right)+
        2\sum_e \frac{\mathbb{D}_{cae}\mathbb{D}_{deb}}{k_aT_{j,a}k_bT_{j,b}k_eT_{j,e}}~.
\end{align}
\end{subequations}
And the integral coefficient $\mathcal{B}_{abcd}$ is given by
\begin{equation}
\label{eq:app:Bcoeff}
    \mathcal{B}_{abcd} = \frac{1}{\Sigma}\iint\dd \Sigma \left(\nabla\chi_a\cdot\nabla\chi_b\right)\left( \nabla\chi_c\cdot\nabla\chi_d\right)~.
\end{equation}

\end{document}